\documentclass[aps,prb,twocolumn,floatfix,amsmath,amssymb]{revtex4-1}

\usepackage{amsmath,bm,amsfonts}
\usepackage[dvipdfmx]{graphicx}

\newfont{\bg}{cmr10 scaled\magstep4}

\newcommand{\bigzerou}{\smash{\lower1.7ex\hbox{\bg 0}}}

\newcommand{\be}{\begin{equation}}
\newcommand{\ee}{\end{equation}}
\newcommand{\bal}{\begin{aligned}}
\newcommand{\eal}{\end{aligned}}

\begin{document}

\title{Methods for constructing parameter-dependent flat band lattices}
  
\author{Toshitaka Ogata, Mitsuaki Kawamura, and Taisuke Ozaki}
\affiliation{
  Institute for Solid State Physics, The University of Tokyo, Kashiwa 277-8581, Japan
}

\date{\today}

\begin{abstract} 
We present two methods for constructing a flat band (FB) system having a flat energy dispersion over the entire Brillouin zone
within tight-binding model, where the resulting Hamiltonian may not be easily obtained by existing methods 
based on a bipartite graph and line graph techniques. 
In the first method, we derive a set of conditions equivalent to the appearance of FBs for a given graph structure. 
This method allows parameter to be tuned so that systems with a small number of sites per unit cell has a FB. 
In the second method, we show that FB systems can be obtained by removing or adding sites to an existing FB system 
under specific rules. 
In particular, the site addition method enables us to construct multiple FB systems stemming from a single FB system. 
The FB system obtained by the second method has the characteristics that the component ratios in the FB eigenstate 
are partially common to the original system. 
We illustrate how lattices having a FB can be constructed by applying the latter method starting from an existing 
lattice such as a kagome lattice, demonstrating that a wide variety of lattices can possess a FB in the band structure. 
\end{abstract}

\maketitle

\section{INTRODUCTION}

Flat band (FB) systems have a band structure with a flat energy dispersion over the entire Brillouin zone. FBs play a central role in a wide variety of phenomena in solids: The ferromagnetism of the Hubbard model with a half-filled FB has been demonstrated for several systems such as Lieb, kagome, and Tasaki lattices \cite{PhysRevLett.62.1201,Mielke_1991,Mielke_1991_2,PhysRevLett.69.1608,Mielke1993}. FB systems with nontrivial topological invariants have also been reported, and the possibility of the room temperature fractional quantum Hall effect has been discussed \cite{PhysRevB.82.075104,Katsura_2010,PhysRevLett.106.236802,PhysRevLett.106.236803,PhysRevLett.106.236804}. The relationship between FB systems and superconductivity has also been discussed.\cite{PhysRevLett.84.143}. Recently, superconductivity has been experimentally confirmed in the twisted bilayer graphene at a specific twist angle, the magic angle, where nearly FBs have been reported by several theoretical calculations \cite{PhysRevLett.99.256802,Bistritzer12233,doi:10.1021/nl902948m,Cao2018,PhysRevX.8.031089}.
It has also become possible to artificially form two-dimensional lattices in cold atom systems \cite{Taiee1500854}, photonic crystal systems \cite{PhysRevLett.114.245504}, and surface systems \cite{Drost2017,Slot2017}. These realizations have triggered interests in the design of FB systems such as Lieb lattice.

Obtaining a variety of FB systems is important for both investigating the phenomena arising from FBs and designing systems with FBs. The conventional way to derive FB systems is to construct a bipartite graph with different size of sublattices \cite{PhysRevB.34.5208}. The FB of lattices like Lieb lattice and dice lattice can be understood from the bipartite graph structure. The FB construction method using line graph structures has also been proposed \cite{Mielke_1991_2,Katsura_2010}. For example, FB for a kagome lattice, which is a line graph of a honeycomb lattice, and a checkerboard lattice, which is a line graph of a square lattice, can be understood from the graph structure.
In addition to those methods, a method of partial line graphs inspired by the line graph method\cite{doi:10.1143/JPSJ.74.1918}, a method of constructing FB systems by combining miniarrays \cite{PhysRevA.94.043831}, and a method of obtaining different systems by transforming a plaquette using a procedure called the origami rule \cite{Dias2015} have been proposed. An attempt to classify FB systems in a 1D system and to generate a system with FBs based on the classification has also been reported \cite{PhysRevB.95.115135,PhysRevB.99.125129}.

Furthermore, a method of constructing FB systems by tuning the parameters rather than the graph structure of the system has been proposed. When a system has a FB, the system corresponding to the power of the Hamiltonian of the system also has a FB. A method to adjust the FB energy using the property has been reported \cite{PhysRevB.99.235118}. FBs that arise when certain conditions are satisfied between parameters are also reported for the distorted Lieb lattice with additional atoms, the bitriangular lattice and the checkerboard lattice with additional atoms \cite{PhysRevB.100.045150}.

In this paper, we propose two methods for constructing FB systems by tuning parameters involved in the tight-binding (TB) model. 
By the methods one can derive a wide range of systems having a FB, which may not be easily obtained by existing methods 
such as a bipartite graph and line graph techniques.
The first method is to tune the parameters for a given system so that the system satisfies a set of conditions for having a FB. The set of conditions is necessary and sufficient for the system to have a FB. Within the set of conditions derived for a given graph structure, one can vary the TB parameters 
with keeping the flatness of a band in the band structure, while the conditions become complicated to handle analytically as the number of sites 
per unit cell increases. The conditions should be satisfied for a given graph structure if the TB system has a FB.
Thus, not only any known FB systems such as Lieb and kagome lattices, but also non-trivial lattices having a FB must 
fulfill the corresponding conditions.
The second method derives a FB system by removing or adding a site, where a FB system is constructed from an existing FB system. 
This method can handle FB systems with a large number of sites per unit cell, and enables us to obtain FB systems with distant TB hopping parameters because additional edges are created in the procedure. 
The site removal and addition procedures can be regarded as an extension and generalization of the method 
by Lee {\it et al.} \cite{PhysRevB.100.045150}, respectively.

The paper is organized as follows: In Sec. 2, we explain how to obtain a set of conditions equivalent to the appearance of FBs and how to tune the parameters to obtain a FB system. In Secs. 3 and 4, we describe how to construct a FB system by removing and adding sites from a known FB system, respectively. 
In Sec. 5, we conclude our methods with future perspectives.

\section{FB conditions in tight-binding (TB) model}

In the spinless TB model, we derive general conditions to have a FB for TB parameters in a system with a given graph structure.
First, let us introduce a lattice with $N$ sites per unit cell, whose TB Hamiltonian is given by 
\be
\label{ham_def}
H=\sum_{i,j}\sum_{\bm{R},\bm{R'}}t_{ji,\bm{R'}}\;\hat{a}^\dag_{j\bm{R}+\bm{R'}}\hat{a}_{i\bm{R}},
\ee
where $i$ and $j$ represent the site index running from $1$ to $N$ in the unit cell, and $\bm{R}$ and $\bm{R'}$ are the lattice vectors specifying 
the position of the unit cell. $\hat{a}^\dag_{i\bm{R}}$ ($\hat{a}_{i\bm{R}}$) is the creation (annihilation) operator for the site $i$ 
in the unit cell specified by the lattice vector $\bm{R}$. $t_{ij,\bm{R'}}$ is the hopping integral from the site $i$ in the unit cell 
of $\bm{R}=\bm{0}$ to the site $j$ in the unit cell of $\bm{R}=\bm{R'}$, while $t_{ii,\bm{0}}$ is the on-site energy 
of the site $i$. The Fourier transformation (FT) of the annihilation operator is defined by 
\be
\label{FT}
\hat{a}_{i\bm{R}}=\frac{1}{\sqrt{M}}\sum_{\bm{k}}e^{-\mathrm{i}\bm{k}\cdot\bm{R}}\hat{c}_{i\bm{k}},
\ee
where $M$ is the total number of unit cells, which is large enough. 
$\hat{c}_{i\bm{k}}$ is an annihilation operator in the reciprocal lattice space. 
The FT of the creation operator can be obtained from the Hermitian conjugate of Eq.~(\ref{FT}). 
Then, using Eq.~(\ref{FT}) and one corresponding to the creation operator the Hamiltonian of Eq.~(\ref{ham_def}) can be rewritten as
\be
\label{e1}
\hat{H}=\sum_{\bm{k}}\sum_{i,j}\left(\sum_{\bm{R}}t_{ji,\bm{R}}\;e^{i\bm{k}\cdot\bm{R}}\right)\hat{c}^\dag_{j\bm{k}}\hat{c}_{i\bm{k}}.
\ee
In the following, we write the $\bm{k}$-dependent Hamiltonian as
\be
\label{e1_2}
H_{ij}(\bm{k})=\sum_{\bm{R}}t_{ij,\bm{R}}\;e^{i\bm{k}\cdot\bm{R}},
\ee
which is nothing but the FT of TB parameter. 

When the system has a FB, the Hamiltonian has an eigenvalue $E_{\rm FB}$ that is independent of the wave vector $\bm{k}$. 
Without losing generality, one can shift the on-site energy of each site by $-E_{\rm FB}$ and set the FB energy to 0. 
Then, the determinant of the Hamiltonian must be zero because the secular equation $|H(\bm{k})-\lambda \bm{I}|=0$ has an eigenvalue of $\lambda=0$ . 
The determinant of the Hamiltonian is explicitly written as
\be
\bal
\label{e2}
|H(\bm{k})|&=\sum_{\sigma\in S_N}\mathrm{sgn}(\sigma)H_{1\sigma(1)}(\bm{k})H_{2\sigma(2)}(\bm{k})\cdots H_{N\sigma(N)}(\bm{k})\\
&=\sum_{\sigma\in S_N}\sum_{\bm{R}}\sum_{\substack{\bm{R}_1,\cdots\bm{R}_N,\\ 
  \bm{R}_1+\cdots+\bm{R}_N=\bm{R}}}\\
&\mathrm{sgn}(\sigma)t_{1\sigma(1),\bm{R}_1}\cdots t_{N\sigma(N),\bm{R}_N}e^{i\bm{k}\cdot\bm{R}},
\eal
\ee
where $S_N$ denotes the symmetric group on a set of $N$ and $\mathrm{sgn}(\sigma)$ denotes the sign of the permutation $\sigma$. 
$\bm{R}_1,\cdots$, and $\bm{R}_N$ run thorough the range satisfying $\bm{R}_1+\cdots+\bm{R}_N=\bm{R}$. 
Since $|H(\bm{k})|=0$ must hold for all $\bm{k}$ if the system has a FB, one can obtain conditions among parameters to have a FB  
by comparing the Fourier coefficients of $e^{\mathrm{i}\bm{k}\cdot\bm{R}}$ as follows:
\be
\bal
\label{e4}
&\sum_{\sigma\in S_N}\sum_{\substack{\bm{R}_1,\cdots,\bm{R}_N \\ \bm{R}_1+\cdots\bm{R}_N=\bm{R}}}\\
&\mathrm{sgn}(\sigma)t_{1\sigma(1),\bm{R}_1}t_{2\sigma(2),\bm{R}_2}\cdots t_{N\sigma(N),\bm{R}_N}=0 \quad (\forall \bm{R}).\\
\eal
\ee
All of these equations must be satisfied in order for the system to have a FB. 
In practice, the range of $\bm{R}$ is limited due to the short range property of TB parameters we generally assume.
Since these equations give a set of conditions equivalent to having a FB for any lattice system, any known FB systems such as Lieb and kagome lattices
should satisfy the FB condition of Eq.~(\ref{e4}) among TB parameters for all $\bm{R}$. However, the application of the method in deriving a FB system 
might be limited to a relatively small system, since the analytic formula derived from Eq.~(\ref{e4}), 
which are given as a pair of $N$-th order equations for $t_{ij,\bm{R}}$, become overly complicated as the number of sites in the unit cell increases. 
Note that the number of non-zero terms in Eq.~(\ref{e4}) scales as $N!(N_{\rm NN}/N)^N$ for a given $\bm{R}$ if the number 
of nearest neighboring sites is $N_{\rm NN}$ on average.

\begin{figure}[t]
\includegraphics[width=8cm]{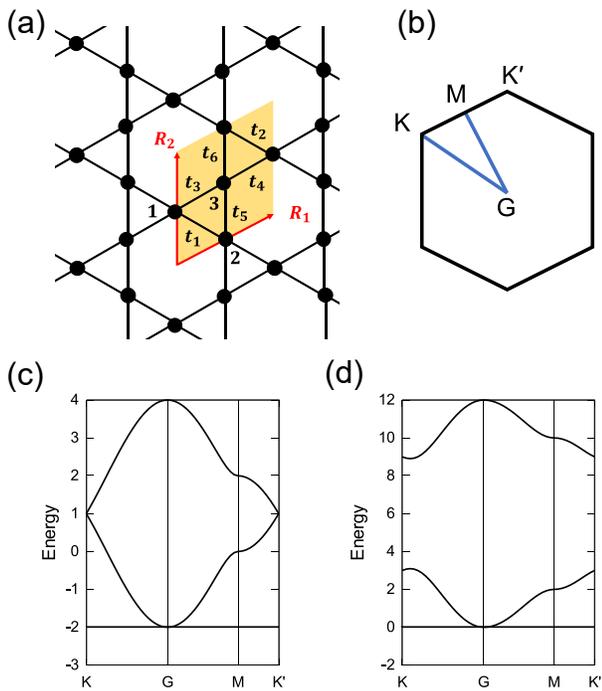}
\caption{(Color Online) (a) Graph structure of a kagome lattice and definition of hopping integrals. (b) Brillouin zone of the kagome lattice.
(c) Band structures of the conventional kagome lattice with parameters of $t_1=t_2=t_3=t_4=t_5=t_6=1$ and $e_1=e_2=e_3=0$.
(d) Band structures of a kagome lattice with tuned parameters given by $t_1=t_2=t_3=t_4=2$, $t_5=t_6=1$, $e_1=8$, and $e_2=e_3=2$.}
\label{kagome}
\end{figure}

As an example, let us investigate the FB conditions for a kagome lattice.
A kagome lattice is a two-dimensional lattice with three sites per unit cell, depicted in Fig.~\ref{kagome}(a). The unit cell is shown in the shaded area with $\bm{R_1}$ and $\bm{R_2}$ as lattice vectors. Assuming that all hopping integrals to the nearest site are $t$ and the onsite energies are zeros, a FB is obtained at $E=-2t$ as shown in Fig.~\ref{kagome}(c), where $t$ is set to be 1. 
Here, we relax the restriction that hopping integrals are the same for all edges and that on-site energies are the same regardless of the sites. The hopping integrals are defined as in Fig.~\ref{kagome}(a), and the on-site energy at the site $i$ is defined as $e_i$. 
So, we have nine variables describing the system, and obtain the following relations among them using Eq.~(\ref{e4}):
\be
\label{e5}
\bal
&e_1e_2e_3+2t_1t_3t_5+2t_2t_4t_6\\
&-e_1(t_5^2+t_6^2)-e_2(t_3^2+t_4^2)-e_3(t_1^2+t_2^2)=0 \quad(\bm{R}=\bm{0}),\\
&t_1t_4t_5+t_2t_3t_6-e_2t_3t_4=0 \quad(\bm{R}=\bm{R}_1,-\bm{R}_1),\\
&t_1t_3t_6+t_2t_4t_5-e_1t_5t_6=0 \quad(\bm{R}=\bm{R}_2,-\bm{R}_2),\\
&t_1t_4t_6+t_2t_3t_5-e_3t_1t_2=0 \quad(\bm{R}=\bm{R}_1-\bm{R}_2,-\bm{R}_1+\bm{R}_2).
\eal
\ee
For the other $\bm{R}$s, all the terms vanish due to the short range property of TB parameters specified in Fig.~\ref{kagome}(a).
Now, we assume that the hopping integrals from $t_1$ to $t_6$ are all non-zero in order not to change 
the graph structure of a kagome lattice. 
In this case, the first equation can be obtained from the other three equations, 
and the FB conditions are written as
\be
\bal
\label{e6}
&e_1=\frac{t_1t_3t_6+t_2t_4t_5}{t_5t_6},\\
&e_2=\frac{t_1t_4t_5+t_2t_3t_6}{t_3t_4},\\
&e_3=\frac{t_1t_4t_6+t_2t_3t_5}{t_1t_2}.
\eal
\ee
From these equations, we see that there are six dimensions of freedom out of the total nine dimensions in which the system has a FB.
For example, when the hopping integrals are non-zero, a FB can be formed for any hopping parameters if the onsite energies are properly tuned using Eq.~(\ref{e6}).
In the case of the kagome lattice, where the hopping integrals to the nearest site are uniformly $1$, Eq.~(\ref{e6}) gives simply $e_1=e_2=e_3=2$. Shifting the energy origin by $-2$ results in the band structure of Fig.~\ref{kagome}(c). As another example, Fig.~\ref{kagome}(d) shows that a FB at energy of $0$ occurs with parameters: $t_1=t_2=t_3=t_4=2$, $t_5=t_6=1$, $e_1=8$, and $e_2=e_3=2$ which satisfy Eq.~(\ref{e6}).

Note that for a kagome lattice case, equivalent equations can be obtained by considering the degrees of freedom of parameters in FB line graph systems because a kagome lattice is a line graph of a honeycomb lattice. However, our method derives the FB conditions more directly, and can be applied to any system, not just line graph systems.

For a given graph structure our method gives the FB condition. So, once a proper graph structure is introduced, 
one can investigate how large extent TB parameters can be varied while keeping the FB conditions of Eq.~(\ref{e4}). 
The evaluation of Eq.~(\ref{e4}) might be useful to study how structural symmetry breakings and atomic substitutions
affect on the FB in known flat band structures. In addition, one might be able to check the degree of coincidence of Eq.~(\ref{e4}) 
if a new material possesses a FB. The procedure will be performed by utilizing TB parameters derived from 
maximally localized Wannier functions (MLWFs) \cite{RevModPhys.84.1419} 
within density functional theory (DFT) \cite{PhysRev.136.B864,PhysRev.140.A1133}.
However, the analytic handling of FB conditions might be limited to relatively small systems, since 
the number of non-zero terms in Eq.~(\ref{e4}) scales as $N!(N_{\rm NN}/N)^N$ as discussed before.
On the other hand, the numerical evaluation of $|H(\bm{k})|$ can be performed by employing matrix factorization techniques
such as LU factorization in O($N^3$) operation.

\section{Construction method of a FB system by site removal}
\subsection{Principle of the construction method}

Apart from the FB conditions defined by Eq.~(\ref{e4}), we propose construction methods of a FB system 
by site removal and addition, which enables us to obtain FB systems with various graph structures, and
to apply to systems with a large number of sites per unit cell.

Let us start with a general case, and consider a system with $N$ sites per unit cell. 
Writing down the Schr\"{o}dinger equation of the system corresponding to a certain eigenvalue $E(\bm{k})$, 
and letting the corresponding eigenstate be $(c_1(\bm{k}),c_2(\bm{k}),\cdots ,c_N(\bm{k}))^T$, we have
\be
\label{e7}
\sum_{j=1}^N H_{ij}(\bm{k})c_j(\bm{k})=E(\bm{k})c_i(\bm{k})  \quad (i=1,2,\cdots,N),
\ee
where the Hamiltonian element $H_{ij}(\bm{k})$ is given by Eq.~(\ref{e1_2}). 
Here, we consider removing a site labeled by $N$ from the system consisting of the $N$ sites 
to generate a system with $(N-1)$ sites per unit cell. 
Defining $\epsilon(\bm{k})=H_{NN}(\bm{k})-E(\bm{k})$, and multiplying $H_{iN}(\bm{k})$ 
by the expression of Eq.~(\ref{e7}) with $i=N$, we obtain the following equation:
\be
\bal
\label{e8}
\sum_{j=1}^{N-1}H_{iN}(\bm{k})H_{Nj}(\bm{k})c_j(\bm{k})+\epsilon(\bm{k}) H_{iN}(\bm{k})c_N(\bm{k})=0 \\
(i=1,2,\cdots,N-1).
\eal
\ee
Next, multiplying the expression of Eq.~(\ref{e7}) for $i=1,2,\cdots,N-1$ by $\epsilon (\bm{k})$, 
and subtracting it from Eq.~(\ref{e8}), we have equations for $(c_1(\bm{k}),c_2(\bm{k}),\cdots ,c_{N-1}(\bm{k}))^T$ 
as follows:
\be
\label{e9}
\bal
\sum_{j=1}^{N-1} (\epsilon(\bm{k})H_{ij}(\bm{k})-H_{iN}(\bm{k})H_{Nj}(\bm{k}))c_j(\bm{k})=E(\bm{k})\epsilon(\bm{k})c_i(\bm{k}) \\
(i=1,2,\cdots,N-1).
\eal
\ee
Here, we assume $\epsilon(\bm{k})\neq 0$ to eliminate a possibility that the eigenstate is localized at the site $N$. Then, for any $\bm{k}$, $(c_1(\bm{k}),c_2(\bm{k}),\cdots ,c_{N-1}(\bm{k}))^T\neq \bm{0}$ holds because if we assumed that $c_1(\bm{k}),c_2(\bm{k}),\cdots ,c_{N-1}(\bm{k})$ are all $0$, then from the equation (\ref{e7}), $c_{N}(\bm{k})=0 $ would hold, which is incompatible with the fact that $(c_1(\bm{k}),c_2(\bm{k}),\cdots ,c_{N}(\bm{k}))^T$ is an eigenstate. 
Equation~(\ref{e9}) shows that the Hamiltonian element $H'_{ij}(\bm{k})=\epsilon(\bm{k})H_{ij}(\bm{k})-H_{iN}(\bm{k})H_{Nj}(\bm{k})$ for the $(N-1)$ sites has the eigenvalue $E(\bm{k})\epsilon (\bm{k})$ and the eigenstate is given by $(c_1(\bm{k}),c_2(\bm{k}),\cdots ,c_{N-1}(\bm{k}))^T$, excluding the normalization factor.

\begin{figure}[t]
\includegraphics[width=8cm]{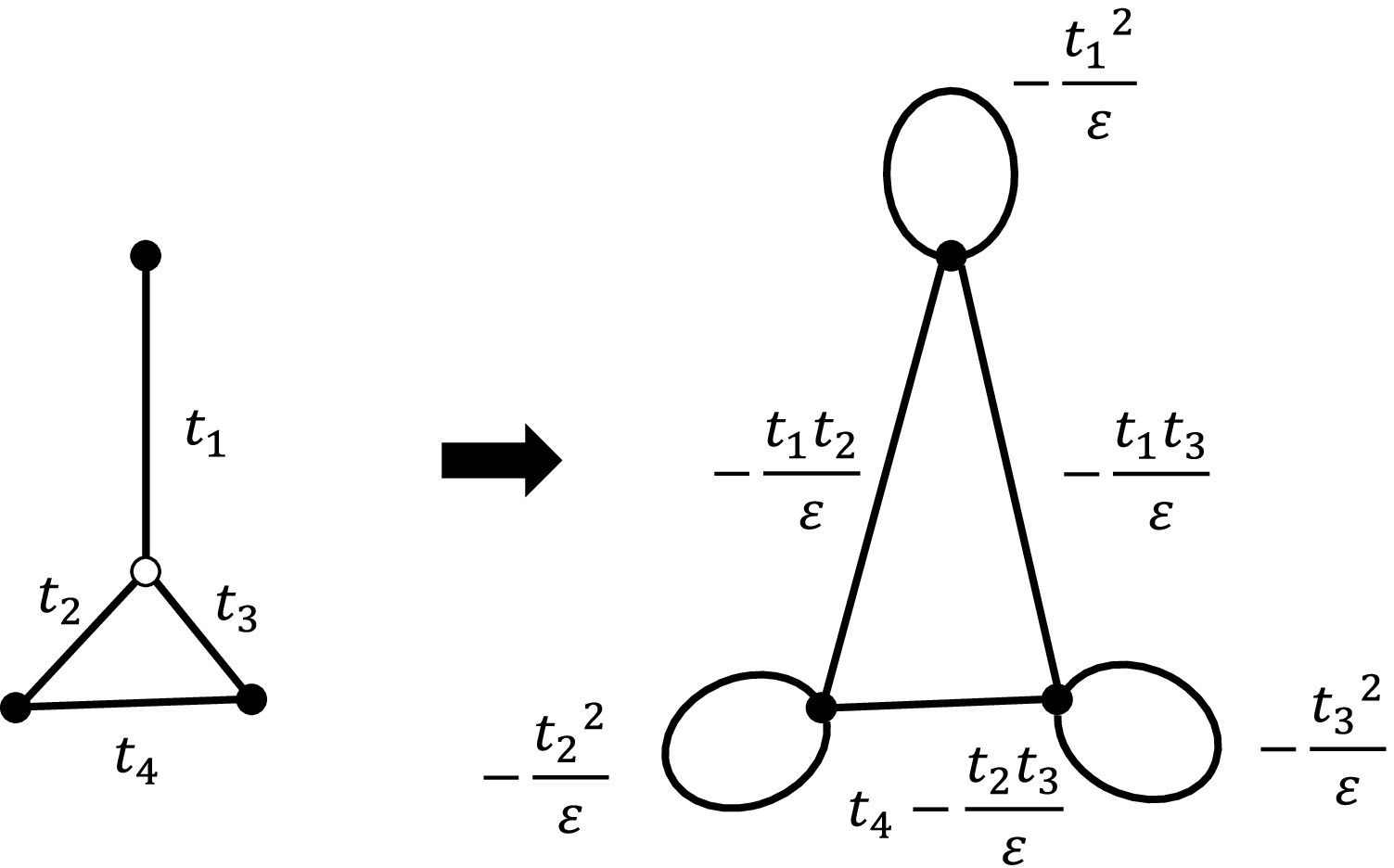}
\caption{A schematic diagram of the site-removal procedure. The FB system on the right lattice is constructed from the FB system on the left lattice. The white circle site is removed, and the procedure creates hoppings between the sites adjacent to the white circle site. $\epsilon$ is defined as the on-site energy of the white circle minus the FB energy.}
\label{site_remove}
\end{figure}

\begin{figure*}[ht]
\includegraphics[width=15cm]{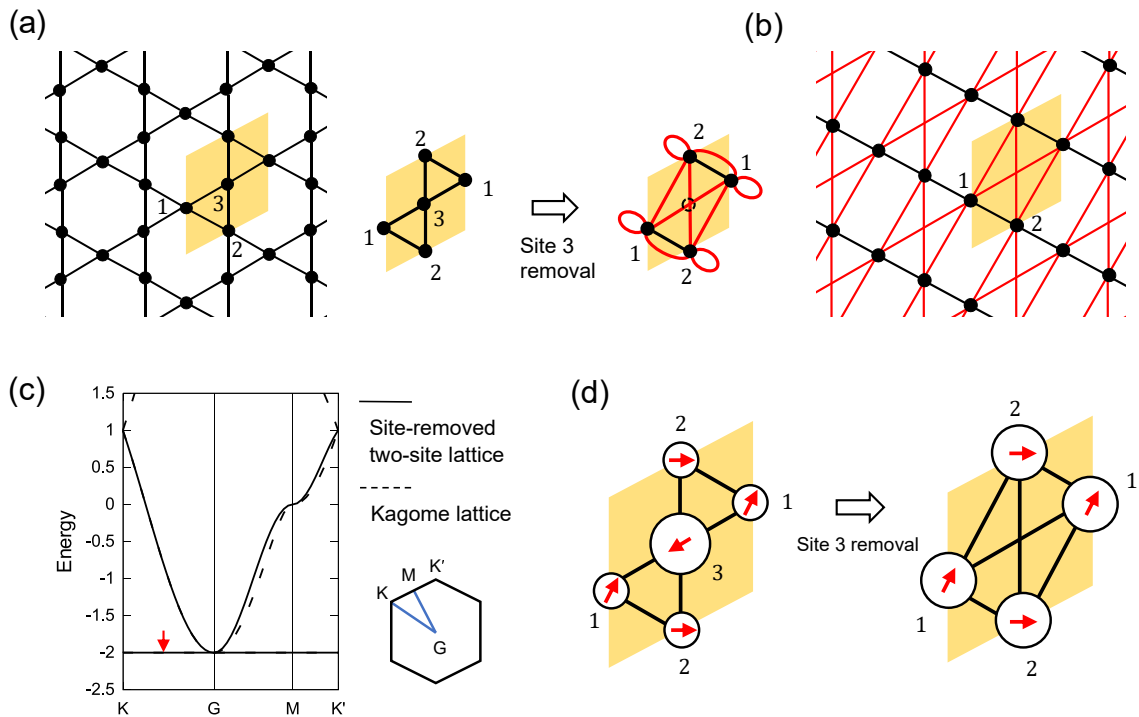}
\caption{(Color Online) (a) Removal procedure of the site $3$ in the kagome lattice with each hopping integral of $1$ and each on-site energy of $0$. The red lines in the right figure are the hoppings of $-1/2$ that are created during the procedure. 
The dashed circle depicts the removed site. (b) A checkerboard type two-site system obtained by the procedure. 
The black and red lines represent hoppings of $1/2$ and $-1/2$, respectively, and the on-site energy of each site is $-1$. 
(c) Band structure before (solid line) and after (dashed line) the site removal procedure, and the Brillouin zone. 
(d) FB eigenstates for the kegome lattice and the obtained two-site lattice at the $\bm{k}$ point indicated by the red arrow in (c), 
where $\bm{k}$ is found to be $(-1.900,1.097)$ if the length of the edge of the rhombus of the unit cell is $1$. 
The radius of the white circle is proportional to the absolute value of the component on each site of the eigenstate, 
and the red arrow in the white circle corresponds to the phase of the component.}
\label{site_remove_kagome}
\end{figure*}

Note that when $\epsilon(\bm{k})=0$, Eq.~(\ref{e8}) becomes
\be
\bal
\label{e10}
\sum_{j=1}^{N-1} H_{iN}(\bm{k})H_{Nj}(\bm{k})c_j(\bm{k})=0 \quad (i=1,2,\cdots,N-1).
\eal
\ee
Equation (\ref{e10}) gives only information about the sites adjacent to the site $N$. 
It depends on the details of the system whether $c_l(\bm{k})\neq 0$ holds 
for at least one site labeled by $l$ adjacent to the site $N$ at each $\bm{k}$. 
If the condition $c_l(\bm{k})\neq 0 $ holds, Eq.~(\ref{e10}) means that the system described 
by the new Hamiltonian $H'_{ij}(\bm{k})=H_{iN}(\bm{k})H_{Nj}(\bm{k})$ consisting of sites 
adjacent to the site $N$ has a FB with zero energy.

Next, let us consider a case where a system with $N$ sites per unit cell has a FB, and 
generate a system consisting of $(N-1)$ sites by applying the method explained above 
to the system with $N$ sites for the FB eigenvalue $E(\bm{k})=E_{\rm FB}$.
In particular, it turns out that $E(\bm{k})\epsilon(\bm{k})$ does not depend on $\bm{k}$
if $H_{NN}(\bm{k})$ does not depend on $\bm{k}$, that is, $\epsilon(\bm{k})$ does not depend on $\bm{k}$. 
Thus, we see that the constructed Hamiltonian also has a FB eigenstate. 
In the following discussion, we consider a case where $\epsilon(\bm{k})=\epsilon\neq 0$.
In this case, Eq.~(\ref{e9}) reads as
\be
\bal
\label{e11}
\sum_{j=1}^{N-1} \left(H_{ij}(\bm{k})-\frac{H_{iN}(\bm{k})H_{Nj}(\bm{k})}{\epsilon}\right)c_j(\bm{k})=E_{\rm FB}c_i(\bm{k})\\
(i=1,2,\cdots,N-1).
\eal
\ee
It is possible to interpret Eq.~(\ref{e11}) as a graphical procedure which generates a system consisting of $(N-1)$ sites 
starting from a system with $N$ sites. A schematic diagram for the graphical procedure is depicted in Fig.~\ref{site_remove}, 
and the detail of the procedure is summarized as follows:
\begin{enumerate}
\item Choose a site labeled by $l$ from the $N$-site system, where $H_{ll}(\bm{k})=e_l\neq E_{\rm FB}$ in order to make $\epsilon(\bm{k})(=H_{ll}(\bm{k})-E_{\rm FB})$ a non-zero constant. For generality we use the label $l$, but it corresponds to the label $N$ in Eq.~(\ref{e7}).   
From Eq.~(\ref{e1_2}), this condition corresponds to $t_{ll,\bm{0}}=e_l$ and $t_{ll,\bm{R}\neq\bm{0}}=0$. In other words, we choose the site $l$, where there is no hopping between the sites $l$ located in different unit cells,
and the on-site energy of the site $l$ is different from $E_{\rm FB}$. 
If such a site does not exist, the site removal procedure cannot be applied to. 
In Fig.~\ref{site_remove}, we choose the white circle site in the left graph as such a site. 

\item Create hoppings between all sites adjacent to the site $l$ chosen by the first step above. 
A hopping of value $-t_{jl,\bm{R}'}t_{li,-\bm{R}}/\epsilon$ is created between the sites $i$ and $j$ which belong to the unit cell $\bm{R}$ and $\bm{R}'$ away, respectively. Here, $\epsilon$ is defined as $\epsilon=e_l-E_{\rm FB}$. 
The same procedure is also applied to the case with $i=j$ and $\bm{R}=\bm{R}'$, and in this case
the on-site energy of the site $i$ in the unit cell $\bm{R}$ away increases by 
$-t_{il,\bm{R}}t_{li,-\bm{R}}/\epsilon=-t_{il,\bm{R}}^2/\epsilon$.
The procedure is schematically depicted in the right graph of Fig.~\ref{site_remove}.

\item Remove the site $l$ and the hoppings between the site $l$ and the other sites.
\end{enumerate}
The resulting $(N-1)$-site system has a FB at energy of $E_{\rm FB}$.

\subsection{Example of the site removal procedure}

The site removal procedure is demonstrated for a kagome lattice as a concrete example.
We choose a kagome lattice with a FB, where the nearest-neighbor hopping integral is uniformly $1$ and the on-site energy of each site is $0$. The left figure in Fig.~\ref{site_remove_kagome}(a) depicts the kagome lattice. 
Let us consider a case of removing the site $3$ in the figure. 
The FB energy of the system is $-2$, which leads to $\epsilon=2$, and the procedure creates a hoppings of $-1/2$, as shown by the red line in the right figure of Fig.~\ref{site_remove_kagome}(a). The removed site is depicted by the dashed circle in the figure. Note that the site $1$ is connected to two removal sites, and the total on-site energy at the sites increases by $-1$, which is the same for the site $2$. The resulting lattice is shown in Fig.~\ref{site_remove_kagome}(b). Here, the on-site energy of the two sites is $-1$, and the hopping integrals depicted by the black and red lines are $1/2$ and $-1/2$, respectively. The graph structure of this lattice is a checkerboard-type lattice. We see that FB kagome lattices and FB checkerboard lattices are associated with the site removal procedure.

Figure~\ref{site_remove_kagome}(c) shows the band structure (solid line) of the system shown in Fig.~\ref{site_remove_kagome}(b)
together with that (dashed line) of the original kagome lattice.
A FB appears at $E_{\rm FB}=-2$, which is the same as the kagome lattice although the other band structure is slightly different from that of the kagome lattice. Figure~\ref{site_remove_kagome}(d) shows that the FB eigenstates at a $\bm{k}$ point indicated by the red arrow in Fig.~\ref{site_remove_kagome}(c) for both the systems, where the radius of the white circle corresponds to the absolute value of the eigenstate component, and the red arrow drawn inside the circle indicates the phase of the component. 
It is confirmed from Fig.~\ref{site_remove_kagome}(d) that the component ratios of the eigenstates of the sites $1$ and $2$ 
do not change during the site removal procedure.

A checkerboard lattice has hoppings between sites $1$ and between sites $2$, so the site removal procedure cannot be applied to any more. Similarly, successive applications of the site removal procedure to any FB system can reduce the system to a system where no further site removal procedure is possible. FB systems can be classified according to the reduced lattice system. 
The consideration of the classification will be in a future work.

\section{Construction method of a FB system by site addition}

\subsection{Principle of the construction method}

In the previous section, we presented the procedure for obtaining a FB system by site removal. 
In this section, we consider the reverse procedure, that is, the procedure for obtaining a $(N+1)$-site FB system 
by adding a site to a $N$-site FB system. 
When the eigenstates corresponding to the FB energy $E_{\rm FB}$ of the $N$-site system are 
$(c_1(\bm{k}),c_2(\bm{k}),\cdots ,c_N(\bm{k}))^T$, the Schr\"{o}dinger equation reads as
\be
\label{e12}
\sum_{j=1}^N H_{ij}(\bm{k})c_j(\bm{k})=E_{\rm FB}c_i(\bm{k}) \quad (i=1,2,\cdots,N).
\ee
Starting from Eq.~(\ref{e12}), let us consider adding a site labeled by $(N+1)$ with the on-site energy of $e_{N+1}\neq E_{\rm FB}$. 
In the site addition procedure, one can freely choose a hopping $H_{iN+1}(\bm{k})=H^*_{N+1i}(\bm{k})$ between the site $(N+1)$ 
and another site $i$. Defining $\epsilon=e_{N+1}-E_{\rm FB}\neq 0$, for each $\bm{k}$ we can always determine $c_{N+1}(\bm{k})$ 
satisfying the following equation:
\be
\label{e13}
\sum_{j=1}^N H_{N+1j}(\bm{k})c_j(\bm{k})+\epsilon c_{N+1}(\bm{k})=0.
\ee
Multiplying Eq.~(\ref{e13}) by $H_{iN+1}(\bm{k})/\epsilon$ for $i=1,2,\cdots,N$ and adding it to Eq.~(\ref{e12}), 
we obtain the Schr\"{o}dinger equation for the Hamiltonian $H'(\bm{k})$ of the $(N+1)$-site system as follows:
\be
\label{e14}
\sum_{j=1}^{N+1} H'_{ij}(\bm{k})c_j(\bm{k})=E_{\rm FB}c_i(\bm{k}) \quad (i=1,2,\cdots,N+1)
\ee
with elements of $H'(\bm{k})$ given by 
\be
\bal
&H'_{ij}(\bm{k})=H_{ij}(\bm{k})+\frac{H_{iN}(\bm{k})H_{Nj}(\bm{k})}{\epsilon} \quad (i,j=1,2,\cdots N),\\
&H'_{iN+1}(\bm{k})=(H'_{N+1i}(\bm{k}))^*=H_{iN+1}(\bm{k}) \quad (i=1,2,\cdots N),\\
&H'_{N+1N+1}(\bm{k})=e_{N+1}.
\eal
\ee
In this case, $H'(\bm{k})$ has a FB of energy $E_{\rm FB}$, and its eigenstates are $(c_1(\bm{k}),c_2(\bm{k}),\cdots ,c_{N+1}(\bm{k}))^T$, excluding the normalization factor.

\begin{figure}[t]
\includegraphics[width=8cm]{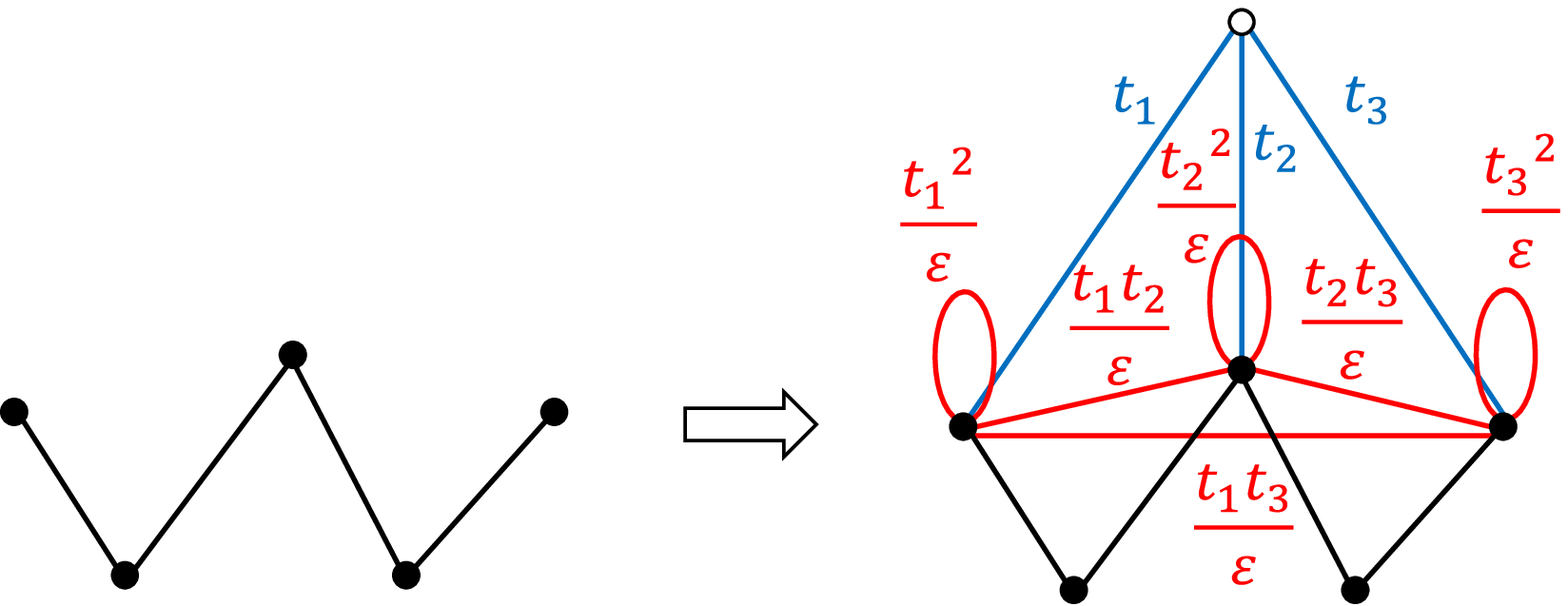}
\caption{(Color Online) A schematic diagram of the site addition procedure. The FB system on the right lattice is constructed from the FB system on the left lattice. The white circle site is added, and the procedure creates hoppings between the sites adjacent to the white circle site. $\epsilon$ is defined as the on-site energy of the white circle minus the FB energy.}
\label{site_add}
\end{figure}

\begin{figure*}[ht]
\includegraphics[width=15cm]{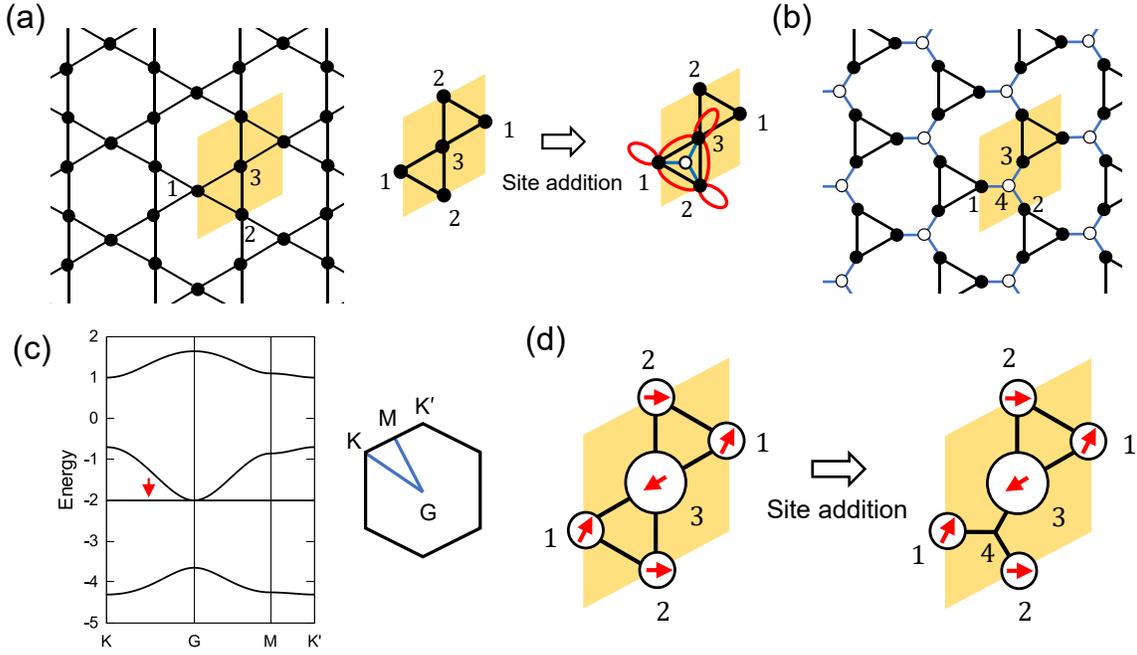}
\caption{(Color Online) (a) Example of the site addition procedure for a kagome lattice. The white circle site is added to the triangular center of the lattice. The blue lines show the hoppings we define, and the red lines show the hoppings created by the procedure. (b) The lattice obtained under parameters where the original hopping and the hoppings created by the site addition procedure are cancelled out. (c) The band structure of the system shown in (b) and and the Brillouin zone, where the hopping integrals are all $1$, and the on-site energies of the white and black sites are $-3$ and $-1$, respectively. (d) The FB eigenstates for the kegome lattice and for the obtained lattice at a $\bm{k}$ point indicated by the red arrow in (c), 
where $\bm{k}$ is found to be $(-1.900,1.097)$ if the length of the edge of the rhombus of the unit cell is $1$.
The radius of the white circle is proportional to the absolute value of the component on each site of the eigenstate, 
and the red arrow in the white circle corresponds to the phase of the component. The site $4$ has no amplitude, 
which is an intrinsic property of the system.}
\label{site_addition1}
\end{figure*}

\begin{figure*}[ht]
\includegraphics[width=15cm]{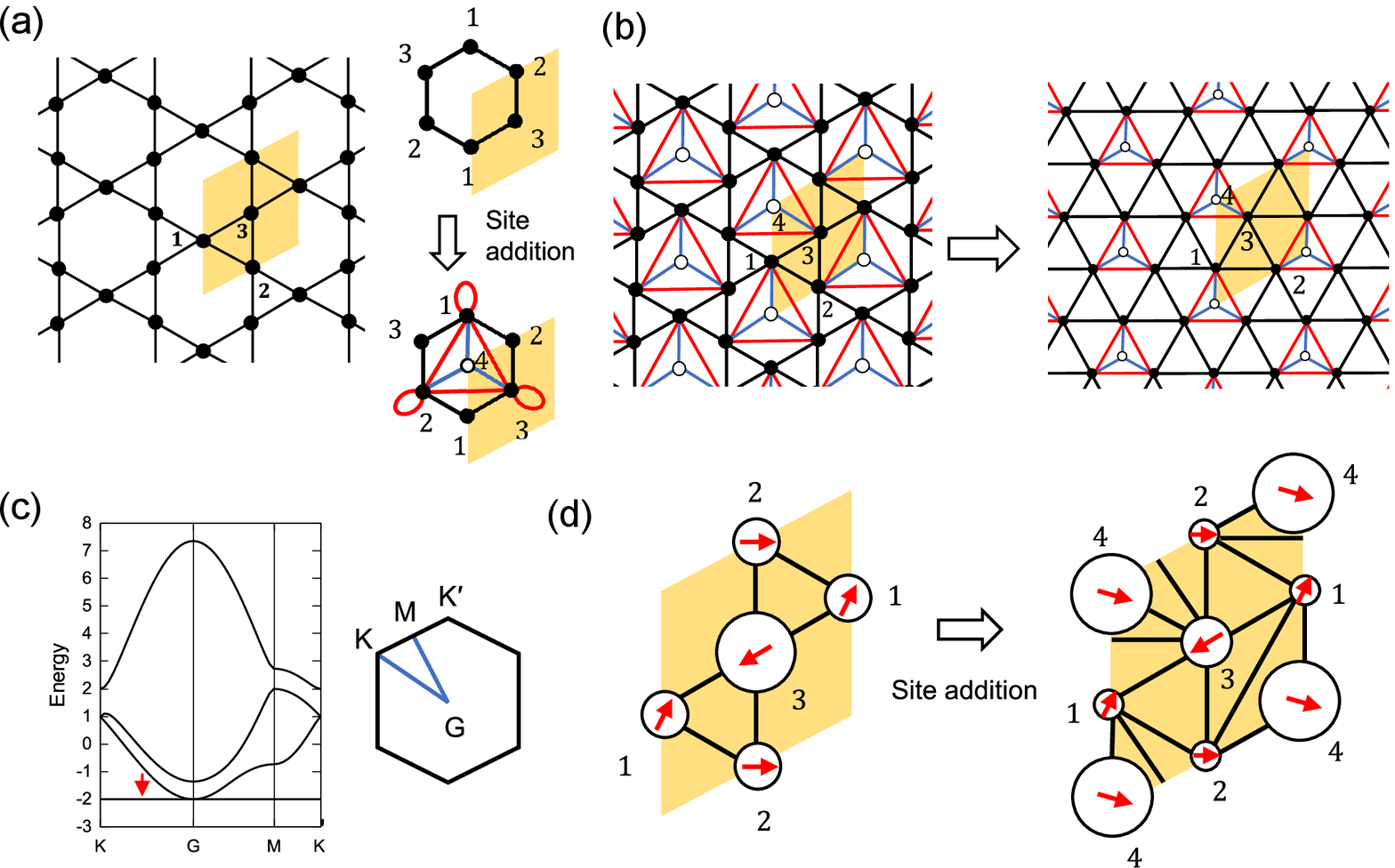}
\caption{(Color Online) (a) Example of the site addition procedure for a kagome lattice. The white circle site is added to the hexagonal center of the lattice. The blue lines show the hoppings we define, and the red lines show the hoppings created by the procedure. (b) The obtained lattice. Transforming the left figure without changing the connection between sites yields the bitriangular lattice shown on the right. (c) The band structure of the system shown in (b) and the Brillouin zone, where the hopping integrals are all $1$, and the on-site energies of the white and black sites are $-1$ and $1$, respectively. (d) The FB eigenstates for the kegome lattice and for the obtained lattice at a $\bm{k}$ point indicated by the red arrow in (c), 
where $\bm{k}$ is found to be $(-1.900,1.097)$ if the length of the edge of the rhombus of the unit cell is $1$. 
The radius of the white circle is proportional to the absolute value of the component on each site of the eigenstate, 
and the red arrow in the white circle corresponds to the phase of the component.}
\label{site_addition2}
\end{figure*}

Similar to the site removal procedure discussed in the previous section, one can interpret Eq.~(\ref{e14})
as a graphical procedure which generates a system consisting of $(N+1)$ sites starting from a system with $N$ sites.
A schematic diagram for the graphical procedure is depicted in Fig.~\ref{site_add}, 
and the detail of the procedure is summarized as follows:
\begin{enumerate}
\item Define an additional site labeled by $(N+1)$ anywhere in the unit cell, and freely choose its on-site energy $e_{N+1}$ 
as long as it is not equal to $E_{\rm FB}$.
The left graph of Fig.~\ref{site_add} corresponds to the original system, and the additional site is depicted as a white circle 
in the right graph.   

\item Define hoppings between the site $(N+1)$ and the sites on the original system. 
The value of hopping integrals can be chosen arbitrary. No hopping between sites $(N+1)$ is assumed in order 
to make $\epsilon(\bm{k})=H_{(N+1)(N+1)}(\bm{k})-E_{\rm FB}$ a non-zero constant, 
that is, we let $t_{(N+1)(N+1),\bm{0}}=e_{(N+1)}$ and $t_{(N+1)(N+1),\bm{R}\neq\bm{0}}=0$ 
for the hopping integrals used in Eq.~(\ref{ham_def}).

\item Create hoppings between all sites adjacent to the site $(N+1)$. 
A hopping of value $+(t_{j(N+1),\bm{R}'}~t_{(N+1)i,-\bm{R}})/\epsilon$ is created 
between sites $i$ and $j$ which belong to the unit cells $\bm{R}$ and $\bm{R}'$ away, respectively. 
The sign is different from that of the site removal procedure. Here, $\epsilon$ is defined as $\epsilon=e_{(N+1)}-E_{\rm FB}$. 
The same procedure is also applied to the case with $i=j$ and $\bm{R}=\bm{R}'$, and in this case
the on-site energy of the site $i$ in the unit cell $\bm{R}$ away increases by 
$+(t_{i(N+1),\bm{R}}~t_{(N+1)i,-\bm{R}})/\epsilon = +t_{i(N+1),\bm{R}}^2/\epsilon$.
The procedure is schematically depicted in the right graph of Fig.~\ref{site_add}.

\end{enumerate}
The resulting $(N+1)$-site system has a FB at energy of $E_{\rm FB}$.

\subsection{Examples of the site addition procedure}

As examples of the site addition procedure, we show two systems in which sites were added to a same kagome lattice 
in different ways. 
Let us introduce a kagome lattice with uniform nearest-neighbor hoppings of $1$ and on-site energies of $0$ 
as shown in the left figure of Fig.~\ref{site_addition1}(a).

First, we consider a case, where sites are added at the triangular center and the hoppings are defined as the blue lines in the right figure of Fig.~\ref{site_addition1}(a). The site addition procedure creates the hoppings of the red lines in the right figure of Fig.~\ref{site_addition1}(a). In particular, when we choose parameters such that the added hoppings cancel the hopping of the original triangle, we obtain a lattice as shown in Fig.~\ref{site_addition1}(b). As an example of the lattice for Fig.~\ref{site_addition1}(b), we choose parameters with on-site energies of $-3$ for the additional sites and hopping integrals of $1$ between the additional sites and the original sites. In this case, the on-site energy at the black circle sites in Fig.~\ref{site_addition1}(b) becomes $-1$. 
Figure~\ref{site_addition1}(c) shows the band structure, where the FB of $E_{\rm FB}=-2$ is retained. Figure~\ref{site_addition1}(d) shows the FB eigenstates at a $\bm{k}$ point indicated by the red arrow in Fig.~\ref{site_addition1}(c) for the original kagome lattice and the obtained lattice. It can be confirmed that the component ratios of the eigenstates on the sites $1$, $2$, and $3$ do not change before and after the site addition procedure. Note that in this system, the eigenstate after the procedure have no amplitude on the added site, which is an intrinsic property of the system.

Next, we consider another case, where sites are added at the hexagonal center and the hoppings between the added site and neighbors are introduced as the blue lines in the right figure of Fig.~\ref{site_addition2}(a). Further, the site addition procedure creates the hoppings of the red lines in the right figure of Fig.~\ref{site_addition2}(a). The left figure of Fig.~\ref{site_addition2}(b) shows the obtained lattice. Transforming this lattice while maintaining the connections between the sites results in the bitriangular lattice shown in the right figure of Fig.~\ref{site_addition2}(b). The value of the hopping integrals shown by the red line are determined by the procedure summarized in the first half of the section. 
Lee {\it et al.} \cite{PhysRevB.100.045150} have derived this relationship between parameters as one condition for a bitriangular lattice to have a FB, from the consideration of TB models of a bitriangular lattice, 
and an experimental realization with Ge atoms of the bitriangular lattice revealed by the site addition procedure has been recently reported \cite{PhysRevB.102.201102}. 

Figure~\ref{site_addition2}(c) shows the band structure when the on-site energy of the added site is $-1$ and the hopping integrals indicated by the blue lines are $1$. In this case, the hopping integrals indicated by the red lines are $1$, and the on-site energies of the black circle sites is $1$. The lattice has the same FB of $E_{\rm FB}=-2$ as the kagome lattice. 
As shown in Fig.~\ref{site_addition2}(d), it can be confirmed that the component ratios of the eigenstates 
on the sites $1$, $2$, and $3$ do not change before and after the site addition procedure.

\section{Conclusion}

In this paper, we have presented the two methods for constructing a FB system, which are a powerful tool to explore unconventional 
structures having a FB.
In the first method, the parameters are tuned using a set of conditions that are necessary and sufficient to produce a FB 
when the system is given. This method can be applied in principle to any system once a proper graph structure is introduced. 
In addition to the important role for derivation of FB conditions, one might be able to utilize the conditions
in order to investigate how large extent TB parameters can be varied while keeping the FB conditions of Eq.~(\ref{e4}).
The evaluation of Eq.~(\ref{e4}) provides a mean to study how structural symmetry breakings and atomic substitutions 
affect on FBs. Furthermore, the degree of coincidence of Eq.~(\ref{e4}) can be checked 
if a novel structure possesses a non-trivial FB, which allows us to unveil novel materials with the FB and the hidden mechanism 
to realize the FB. By employing TB parameters derived from MLWFs \cite{RevModPhys.84.1419} 
within DFT \cite{PhysRev.136.B864,PhysRev.140.A1133}, one can numerically evaluate Eq.~(\ref{e4}).
It seems to be interesting to analyze peculiar FBs found in materials with relatively complicated structures 
such as $R$Co$_5$ ($R$=rare earth) compounds \cite{PhysRevB.91.165137} 
and superstructures of Ag atoms on Si(111) surface \cite{PhysRevB.64.205316,SurfaceScience.600.3141,PhysRevB.77.235425}.
It is also noted that the analytic handling of FB conditions might be limited to relatively small systems, 
since the number of non-zero terms in Eq.~(\ref{e4}) scales as $N!(N_{\rm NN}/N)^N$.
On the other hand, the numerically evaluation of $|H(\bm{k})|$ can be performed in O($N^3$) operation 
by using matrix factorization techniques such as LU factorization . 

The second method provides the construction procedure for generating a series of FB systems by site addition/removal. 
In this method, an existing FB system is manipulated to construct a FB system with a different number of sites per unit cell by one. 
The method by site removal has degrees of freedom in selecting the removal site. 
The method by site addition has degrees of freedom in defining the hoppings between the additional site and 
the original sites and in choosing the on-site energy of the additional site. 
These degrees of freedom produce a wide range of FB systems. 
It is also possible to relate different FB systems by these methods.
The site removal and addition procedures can be regarded as an extension and generalization of the method 
by Lee {\it et al.} \cite{PhysRevB.100.045150}, respectively. 

Although we have discussed two-dimensional systems as illustrations of the two methods, both the methods can be applied to 
three-dimensional systems without any modification. It is also noted that the two methods are applicable even 
for the case that each site has multiple orbitals, which is likely to occur in real materials. 
In this case, both the methods are applied to systems with multiple orbitals 
on a single site by treating each orbital as a single site.

It will be in future studies to explore the origin of the parameter-dependent FB systems by using the variety of FB systems 
obtained by these construction methods, and to understand the obtained systems within the framework of the graph theory.
Exploration of real materials corresponding to lattices with a FB designed by the proposed methods is also an interesting future direction.

\begin{acknowledgments}
We would like to acknowledge Dr. Chi-Cheng Lee for stimulating discussion on FB, which motivated us to study the issue, and 
Prof. Hosho Katsura and Prof. Isao Maruyama for their instructive Japanese review article on FB, which allowed us to 
get familiar with the field. This work was partly supported by JSPS KAKENHI Grant Number 20H00328.

\end{acknowledgments}

\newcommand{\journal}[4]{#1 {\bf #2}, #3 (#4).}
\newcommand{\PR}{Phys. Rev.}
\newcommand{\PRB}{Phys. Rev. B}
\newcommand{\PRX}{Phys. Rev. X}
\newcommand{\PRL}{Phys. Rev. Lett.}
\newcommand{\JPA}{J. Phys. A}
\newcommand{\JPSJ}{J. Phys. Soc. Jpn.}
\newcommand{\RMP}{Rev. Mod. Phys.}
\newcommand{\SurfSci}{Surf. Sci.}

\end{document}